\documentclass{article}

\usepackage[final,nonatbib]{neurips_2019_ml4ps}

\usepackage[utf8]{inputenc} 
\usepackage[T1]{fontenc}    
\usepackage{hyperref}       
\usepackage{url}            
\usepackage{booktabs}       
\usepackage{amsfonts,amsmath}       
\usepackage{nicefrac}       
\usepackage{microtype}      
\usepackage{graphicx}
\usepackage{subfig}

\usepackage{color}

\newcommand{\St}{\mathcal{S}}
\newcommand{\T}{\mathcal{T}}
\newcommand{\A}{\mathcal{A}}

\title{Value-Added Chemical Discovery Using Reinforcement Learning}

\author{%
  Peihong Jiang \\
  Brown University \\
  \texttt{peihong\_jiang@brown.edu}
  \And
  Hieu Doan\\
  Materials Science Division \\
  Argonne National Laboratory\\
  Lemont, IL 60439 \\
  \texttt{hadoan@anl.gov} \\
  \And
  Sandeep Madireddy  \\
  Mathematics and Computer Science Division \\
  Argonne National Laboratory\\
  Lemont, IL 60439 \\
  \texttt{smadireddy@anl.gov} \\
  \And
  Rajeev Surendran Assary  \\
  Materials Science Division \\
  Argonne National Laboratory\\
  Lemont, IL 60439 \\
  \texttt{assary@anl.gov} \\
  \And
  Prasanna Balaprakash  \\
  Mathematics and Computer Science Division \& \\
  Leadership Computing Facility\\
  Argonne National Laboratory\\
  Lemont, IL 60439 \\
  \texttt{pbalapra@anl.gov} \\
}

\begin{document}

\maketitle

\begin{abstract}
 Computer-assisted synthesis planning aims to help chemists find better reaction pathways faster. Finding viable and short pathways from sugar molecules to value-added chemicals can be modeled as a retrosynthesis planning problem with a catalyst allowed. This is a crucial step in efficient biomass conversion. The traditional computational chemistry approach to identifying possible reaction pathways involves computing the reaction energies of hundreds of intermediates, which is a critical bottleneck in silico reaction discovery. Deep reinforcement learning has shown in other domains that a well-trained agent with little or no prior human knowledge can surpass human performance. While some effort has been made to adapt machine learning techniques to the retrosynthesis planning problem, value-added chemical discovery presents unique challenges. Specifically, the reaction can occur in several different sites in a molecule, a subtle case that has never been treated in previous works. With a more versatile formulation of the problem as a Markov decision process, we address the problem using deep reinforcement learning techniques and present promising preliminary results.

\end{abstract}

\section{Introduction}
Chemical transformation is the basis of every aspect of industrial processes including the production of drugs, chemicals, and transportation fuels. Artificial intelligence---in particular, machine learning (ML)---and improved materials understanding present a unique opportunity to provide design rules for utilizing easily accessible carbon reserves in the world by transforming them to value-added chemicals. In order to enable and maximize these chemical transformations, a detailed understanding of the mechanistic steps and knowledge of shortest viable discovery pathways are essential.  
Existing discovery approaches, however, are either manually driven or based on trial and error. Automatically discovering transformation pathways by using ML has the potential to revolutionize and accelerate the  discovery of chemicals and novel reaction pathways. 

\begin{figure}[htp] 
    \centering
    \subfloat[Acid-catalyzed aqueous chemistry: fructose to HMF]{%
        \includegraphics[width=0.5\textwidth]{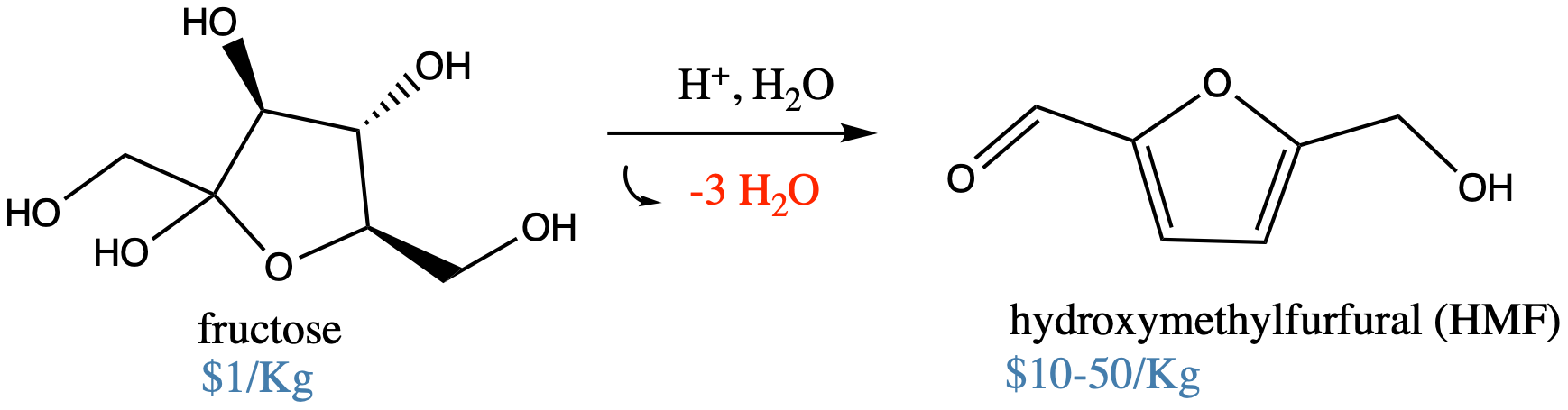}%
        \label{fig:scheme}%
        }%
   \hspace{1em}
    \subfloat[Protonating fructose at different reaction centers leads to distinct offspring]{%
        \includegraphics[width=0.4\textwidth]{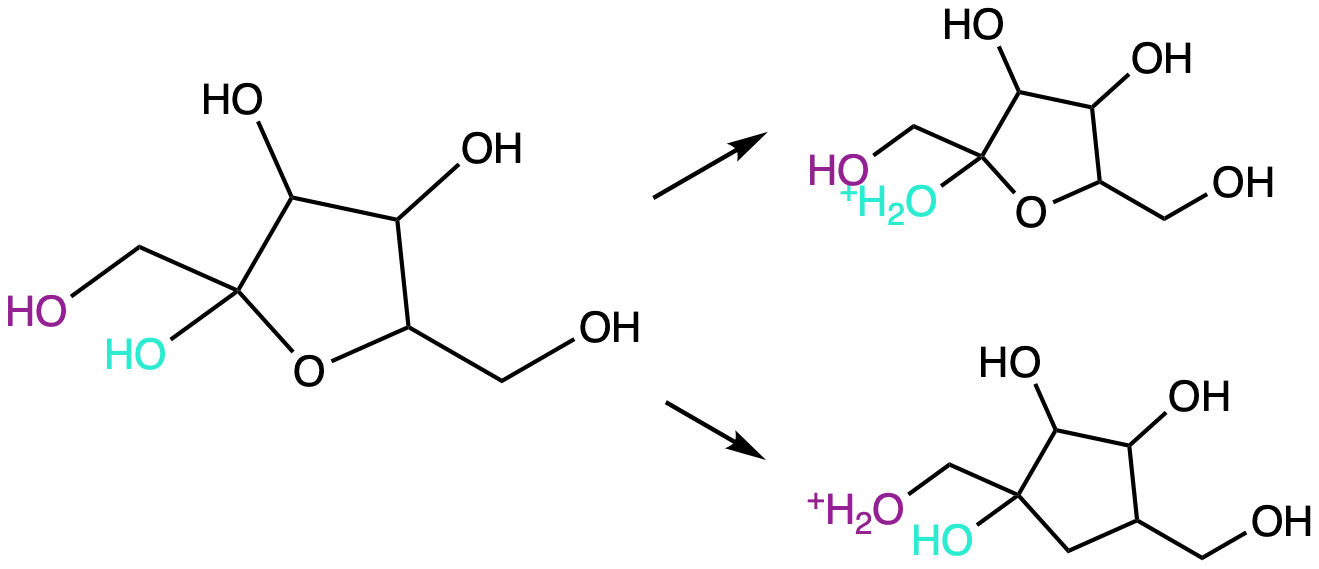}%
        \label{fig:offspring}%
        }%
    \caption{Illustration of chemical reactions on the fructose molecule.}
\end{figure}

Through various chemical transformations  carbon, oxygen, and hydrogen atoms of biomass can be utilized to form useful $\mathrm{C_xH_yO_z}$ candidates. In this regard, automated data-driven adaptive algorithms can play a crucial part in optimizing the desired pathways for the production of novel compounds or identifying viable and cost-effective synthetic routes. To demonstrate this, we have chosen an example of aqueous acid-catalyzed conversion of a fructose molecule to a value-added compound, hydroxy methyl furfural (HMF). This transformation is equivalent to three consecutive dehydration reactions (removal of water molecule) from fructose (shown in Figure \ref{fig:scheme}). We have developed an automatic reaction pathway generator (in Python) based on chemistry rules. This code utilizes  RDKit \cite{landrum2006rdkit}, an open-source cheminformatics software kit that includes an implementation of chemical reactions based on the SMILES arbitrary target specification (SMARTS). We have postulated rules for reactions associated with the carbohydrate chemistry: (a) protonation/deprotonation, (b) dehydration, (c) hydride shift, (d) ring opening (C-O bond cleavage) upon protonation, (e) ring closure (C-O bond) formation upon protonation, (f) ring contraction/expansion (5-6-7 membered rings), (g) keto-enol transformation, (h) addition of water on keto group to form diols, and (i)  formation of formic acid from terminal diols.

Notable recent ML approaches for molecular structure design with sequential chemical transformation stem from the work of Segler et al. \cite{3nmcts} and Coley et al. \cite{coley}, where a reaction template-based Monte Carlo tree search approach and graph convolutional neural-network-based supervised learning approach are adopted, respectively. Schrek et al. \cite{schreck} recently used deep reinforcement learning  to determine optimal reaction paths, an approach that has great potential for synthesis of unfamiliar molecules. However, all previous works   implicitly assume that there is one reaction center possible within a molecule given a particular action template, or they never explicitly state how they handle the multireaction center case. Furthermore, both \cite{3nmcts} and \cite{schreck} mentioned that the quality of the reaction template or the choice of the template is one of the major reasons for the success. Unfortunately, a molecule with several reaction centers given a single reaction template in the sense of \cite{3nmcts, schreck, coley} is ubiquitous in our scenario,  indicating that the crux of our problem is fundamentally different from what the previous works were able to address. Even with computational heavy quantum chemistry methods, 
one can  have a not-quite-accurate estimation of which reaction center has the best chance. This situation motivates us to construct a more versatile formulation of the problem as a  Markov decision process, in the hope that the agent will be able to implicitly learn the underlying probability distribution of the reaction centers through self-play.

\section{Reinforcement learning for chemical synthesis}
We formulate the chemical synthesis problem as a Markov decision process (MDP) \cite{suttonbarto} to make it amenable to the use of reinforcement learning techniques.
 An MDP is a tuple ($\mathcal{S}$, $\mathcal{A}$, $\mathcal{T}$, $r$), where $\St$ denotes the state space, $\A$ the action space, $\T(s,a,s')$ the transition model, and $r$ the reward function $r:\St \times \A \rightarrow \mathbb{R}$, respectively. In our study, a state $s \in \St$ is a set of molecules, and an action $a\in \A (s)$ is one of the reactions from (a)--(i) introduced in the preceding section. We chose to have the action space vary with the states because although  only one SMART template represents each type of reaction, the actual reaction can happen at any site that abides by the chemistry rules. For example, a fructose molecule has six distinct sites (hydroxyl group) where protonation can happen, with different probability determined by the molecule structure and thermodynamics property, which is shown  in \cite{rajeev} to be an important factor. Previous works, including those of Segler et al. \cite{3nmcts} and Schrek et al. \cite{schreck}, classify actions only up to the reaction center. The major concern in both papers was choosing the most likely reaction template at each state. While such simplification is appropriate for most chemical reactions, capturing the underlying probability distribution is simply  not enough when several possible sites are present, as is crucial in our scenario. As a concrete example, a fructose molecule and two distinct reactive offspring from protonation are shown in Figure \ref{fig:offspring}. Both \cite{3nmcts} and \cite{schreck} would represent these as a single action, whereas our formulation distinguishes the reaction centers. 

The agent interacts with this environment by choosing a sequence of actions starting from the initial state and receives a positive reward if the goal state is reached within the maximum steps allowed. Otherwise a negative reward is used to penalize the choices made. The goal of the agent is then to learn an optimal policy function in order to maximize the rewards.

\section{Experiments}

Our approach starts by reading the SMILES string of the parent (fructose) and  applying the protonation rule. Doing so is equivalent to the first step of the acid-catalyzed reaction. For example, fructose has six oxygen atoms; therefore, fructose would have six unique reaction centers and protonation results in the formation of six reactive offspring from fructose. Starting from three initial reactants (fructose, water, and proton), all reaction rules are applied to each reactant one by one. As product species are generated, they are added to the current reactant pool if they are not already in it. The process propagates until no new product can be formed or the reactant list cannot  be updated any further. If we  account only for the products with oxidation s1 or less (neutral),  2,500 reactions  can be generated from the initial three reactants. 
The initial and goal state have been tailored to this data set. We use fructose as initial state and HMF as a goal state. 
We note that most of the reactions in the generated data set are reversible. By reversing the actions, the characterization of the data set changes, allowing us to test the agent's ability to generalize. After reversing the reactions, we run an experiment with fructose as the  goal state and HMF as initial state, respectively. 

Both the original data set and its reversed variant are manually generated. They are meant to simulate the environment on a smaller scale for validation before the full-scale study. Eventually, the available actions $\A(s)$ will be generated by determining the possible sites given the current molecule by the algorithm in an ad hoc fashion. 

The molecules are represented by using a Morgan fingerprint folded to 1,000 dimensions, prepared by using RDKit \cite{landrum2006rdkit}. The Morgan fingerprint is shown to be similar to ECFP4 in most cases according to the description in the online documentation of RDKit \cite{landrum2006rdkit}. To set up the experiment, we implement an OpenAI gym \cite{gym} environment, and we train the policy network with the Proximal Policy Optimization (PPO) algorithm \cite{PPO}. The policy network is modeled by a 128-unit LSTM network \cite{LSTM}. We believe the choice of policy network and training algorithm is appropriate for the following reasons. If we treat our environment as a graph, with states as nodes and actions as edges, or pairs of reactions, for example protonation/deprotonation,  potential loops can be created. Even longer loops are feasible. We would like the agent's policy network to remember the actions it had taken before; and, by receiving negative rewards, the agent would learn to avoid such loops. As for using PPO to update the policy, we note that rewards are given only at the end of each trajectory.

PPO is known to work well on sparse reward problems \cite{OpenAI_dota}. Compared with other policy gradient algorithms such as TRPO \cite{TRPO} or DDPG \cite{DDPG}, it is easily scalable, is more robust, and needs little hyperparameter tuning \cite{PPO}. 

A trajectory is defined as an attempted path from the initial state to a goal state. Rewards are received only once in each trajectory. Because of the nature of chemical reactions, it is impractical to consider synthesis paths longer than $M=20$ steps. 
Let $t$ denote the number of steps. A trajectory has three possible outcomes: (1) the agent reaches goal state at $t = T\leq M$ steps and receives a reward of $1+1/T$; (2) the agent reaches a dead-end state, where no more reactions can happen, and receives a reward of $-1$, or (3) the agent does reach goal state or dead-end state within $M$ steps and receives a reward of $-1$.  

\section{Results}

\begin{figure}[htp] 
    \centering
    \subfloat[fructose to HMF]{%
        \includegraphics[width=0.5\textwidth]{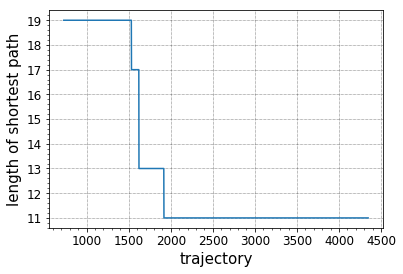}%
        \label{fig:fructose}%
        }%
    \hfill%
    \subfloat[HMF to fructose]{%
        \includegraphics[width=0.5\textwidth]{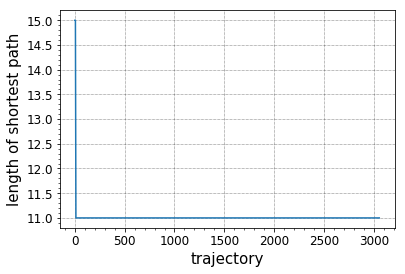}%

        \label{fig:hmf}%
        }%
    \caption{Length of shortest paths with different start states.}
\end{figure}
In Figures \ref{fig:fructose} and \ref{fig:hmf}, the number of steps of shortest paths our agent found was plotted against the number of trajectories. The initial impression is that the agent does gradually learn the shortest paths, but the learning experience varies depending on the start state. The forward direction (fructose to HMF) converges more slowly, possibly because at many states there are more choices of action. During our experiments we found that the performance of the agent is consistent every time we retrain the policy network; therefore no average is taken. 

\begin{figure}
    \centering
    \includegraphics[width=0.95\textwidth]{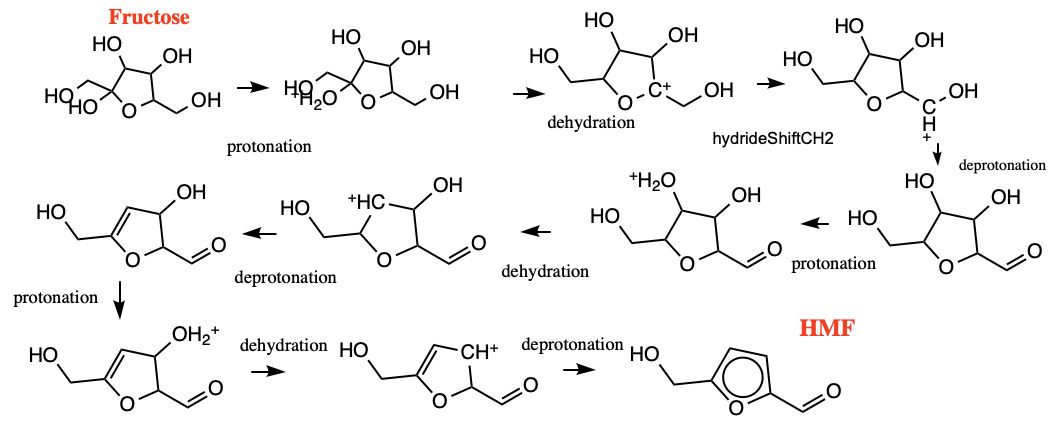}
    \caption{One shortest fructose-HMF reaction sequence identified by reinforcement learning.}
    \label{fig:sample_path}
\end{figure}

Our experiment, although simple, has demonstrated great potential. Data exploration shows that the test data set and its reverse have varied characteristics in terms of the maximum number of actions available across all states and the umber of dead-end states. The agent performed well nonetheless in both cases. Moreover, the agent has little knowledge about the underlying chemistry other than that all molecules are represented by Morgan fingerprints. Not only is the human factor  removed from the discovery process, but the training overhead found in \cite{3nmcts} is also avoided. The simplicity does not mean the performance is compromised. In fact, one of the shortest paths found in Figure \ref{fig:sample_path} is identical to one of the shortest paths identified by chemists in \cite{rajeev}. To compute the thermodynamic landscape as in \cite{rajeev} is in general difficult. The agent has demonstrated the ability to learn through self-play, which is extremely helpful in our scenario.

\section{Discussion}

This work is only the beginning of an exciting project. We  point out some future directions that are  worth exploring. Our next step is to assess how this  approach generalizes to other initial/goal state configurations as well as to larger data sets. We are developing methods to compute possible actions at each state instead of generating them manually, which would become infeasible with various different start states.  The molecule structure largely determines the likelihood of the site at which a particular reaction is going to happen. By converting to Morgan fingerprints, however, this structural information is partially lost in the translation. Therefore, we are planning to change the representation of molecules to graphs so that the agent may be able to learn more directly from the representation rather than using Morgan fingerprints.  Following up the previous idea, we hypothesize that a pretrained network for predicting sites, ideally working directly with the graph representation, will help the agent learn faster in a larger data set. 

\section*{Acknowledgments}
This material is based upon work supported by the U.S.\ Department of Energy 
(DOE), Office of Science, Office of Advanced Scientific Computing Research, under
Contract DE-AC02-06CH11357. We gratefully acknowledge the computing resources provided and operated by the Joint Laboratory for System Evaluation (JLSE) at Argonne National Laboratory. This work was conducted in part by the Computational Chemistry Physics Consortium (CCPC), which is supported by the Bioenergy Technologies Office (BETO) of Energy Efficiency \& Renewable Energy (EERE). P. Jiang was funded by NSF through the MSGI program during her time at Argonne National Laboratory.

\bibliographystyle{plain}
\bibliography{main}

\begin{thebibliography}{10}

\bibitem{rajeev}
Rajeev Assary, Taejin Kim, John Low, Jeff Greeley, and Larry A~Curtiss.
\newblock Glucose and fructose to platform chemicals: Understanding the
  thermodynamic landscapes of acid-catalysed reactions using high-level ab
  initio methods.
\newblock {\em Physical Chemistry Chemical Physics : PCCP}, 14, 08 2012.

\bibitem{gym}
Greg Brockman, Vicki Cheung, Ludwig Pettersson, Jonas Schneider, John Schulman,
  Jie Tang, and Wojciech Zaremba.
\newblock {OpenAI Gym}, 2016.

\bibitem{coley}
Connor~W. Coley, Wengong Jin, Luke Rogers, Timothy~F. Jamison, Tommi~S.
  Jaakkola, William~H. Green, Regina Barzilay, and Klavs~F. Jensen.
\newblock A graph-convolutional neural network model for the prediction of
  chemical reactivity.
\newblock {\em Chem. Sci.}, 10:370--377, 2019.

\bibitem{LSTM}
Sepp Hochreiter and Jürgen Schmidhuber.
\newblock Long short-term memory.
\newblock {\em Neural Computation}, 9(8):1735--1780, 1997.

\bibitem{landrum2006rdkit}
Greg Landrum et~al.
\newblock {RDKit}: Open-source cheminformatics, 2006.

\bibitem{DDPG}
Timothy~P. Lillicrap, Jonathan~J. Hunt, Alexander Pritzel, Nicolas Manfred~Otto
  Heess, Tom Erez, Yuval Tassa, David Silver, and Daan Wierstra.
\newblock Continuous control with deep reinforcement learning.
\newblock {\em CoRR}, abs/1509.02971, 2015.

\bibitem{OpenAI_dota}
OpenAI.
\newblock {OpenAI Five}.
\newblock \url{https://blog.openai.com/openai-five/}.

\bibitem{schreck}
John~S. Schreck, Connor~W. Coley, and Kyle J.~M. Bishop.
\newblock Learning retrosynthetic planning through simulated experience.
\newblock {\em ACS Central Science}, 5(6):970--981, 06 2019.

\bibitem{TRPO}
John Schulman, Sergey Levine, Pieter Abbeel, Michael Jordan, and Philipp
  Moritz.
\newblock Trust region policy optimization.
\newblock In Francis Bach and David Blei, editors, {\em Proceedings of the 32nd
  International Conference on Machine Learning}, volume~37 of {\em Proceedings
  of Machine Learning Research}, pages 1889--1897, Lille, France, 07--09 Jul
  2015. PMLR.

\bibitem{PPO}
John Schulman, Filip Wolski, Prafulla Dhariwal, Alec Radford, and Oleg Klimov.
\newblock Proximal policy optimization algorithms.
\newblock {\em ArXiv}, abs/1707.06347, 2017.

\bibitem{3nmcts}
Marwin H.~S. Segler, Mike Preuss, and Mark~P. Waller.
\newblock Planning chemical syntheses with deep neural networks and symbolic
  {AI}.
\newblock {\em Nature}, 555:604 EP --, 03 2018.

\bibitem{suttonbarto}
Richard~S. Sutton and Andrew~G. Barto.
\newblock {\em Introduction to Reinforcement Learning}.
\newblock MIT Press, Cambridge, MA, USA, 1998.

\end{thebibliography}

\begin{center}
    \framebox{\parbox{5in}{
    The submitted manuscript has been created by UChicago Argonne, LLC, Operator of Argonne National Laboratory (``Argonne''). Argonne, a U.S. Department of Energy Office of Science laboratory, is operated under Contract No. DE-AC02-06CH11357. The U.S. Government retains for itself, and others acting on its behalf, a paid-up nonexclusive, irrevocable worldwide license in said article to reproduce, prepare derivative works, distribute copies to the public, and perform publicly and display publicly, by or on behalf of the Government. The Department of Energy will provide public access to these results of federally sponsored research in accordance with the DOE Public Access Plan. \url{http://energy.gov/downloads/doe-public-access-plan}}}
    \normalsize
\end{center}

\end{document}